\documentclass[fleqn,twoside,onecolumn,nofootinbib,showkeys,11pt]{revtex4}% 
\usepackage{verbatim}
\usepackage[pdftex,plainpages=false]{hyperref}
\usepackage{amsmath}
\usepackage{array}
\usepackage{cleveref}%
\usepackage{cmap} %
\usepackage[utf8]{inputenc} %
\usepackage[english]{babel}
\usepackage[T2A]{fontenc}
\usepackage{amstext}
\usepackage{amssymb}
\usepackage{graphicx}
\usepackage{enumitem}
\usepackage{booktabs}
\usepackage{bm}

\begin{document}
\title{Possibilities of the X-ray Diffraction Data Processing Method for Detecting Reflections with Intensity Below the Background Noise Component}
%{POSSIBILITIES OF THE X-RAY DIFFRACTION DATA PROCESSING METHOD FOR DETECTING REFLECTIONS WITH INTENSITY BELOW THE BACKGROUND NOISE COMPONENT}%

\author{S.\,V.\,Gabielkov}%1
\affiliation{Institute for Safety Problems of Nuclear Power Plants, Nat. Acad. of Sci. of Ukraine}%1
\address{36a, Kirova str., Chornobyl 07270, Ukraine}%1
\email{s.gabelkov@ispnpp.kiev.ua} %e-mail 1

\author{I.\,V.\,Zhyganiuk}%1
\affiliation{Institute for Safety Problems of Nuclear Power Plants,   Nat. Acad. of Sci. of Ukraine}%1
\address{36a, Kirova str., Chornobyl 07270, Ukraine}%1
\email{ }%e-mail 1

\author{A.\,D.\,Skorbun}%1
\affiliation{Institute for Safety Problems of Nuclear Power Plants,   Nat. Acad. of Sci. of Ukraine}%1
\address{36a, Kirova str., Chornobyl 07270, Ukraine}%1
\email{ }%e-mail 1

\author{V.\,G.\,Kudlai}%1
\affiliation{Institute for Safety Problems of Nuclear Power Plants,   Nat. Acad. of Sci. of Ukraine}%1
\address{36a, Kirova str., Chornobyl 07270, Ukraine}%1
\email{ }%e-mail 1

\author{B.\,S.\,Savchenko}%1
\affiliation{Institute for Safety Problems of Nuclear Power Plants,   Nat. Acad. of Sci. of Ukraine}%1
\address{36a, Kirova str., Chornobyl 07270, Ukraine}%1
\email{ }%e-mail 1

\author{P.\,E.\,Parkhomchuk}%1
\affiliation{Institute for Safety Problems of Nuclear Power Plants,   Nat. Acad. of Sci. of Ukraine}%1
\address{36a, Kirova str., Chornobyl 07270, Ukraine}%1
\email{ }%e-mail 1

\author{S.\,O.\,Chikolovets}%1
\affiliation{Institute for Safety Problems of Nuclear Power Plants,   Nat. Acad. of Sci. of Ukraine}%1
\address{36a, Kirova str., Chornobyl 07270, Ukraine}%1
\email{ }%e-mail 1

\begin{center}
  {\small
  \textbf{Published as:}\\[2pt]
  S.\,V.~Gabielkov, I.\,V.~Zhyganiuk, A.\,D.~Skorbun, V.\,G.~Kudlai,
  B.\,S.~Savchenko, P.\,E.~Parkhomchuk, and S.\,O.~Chikolovets,\\
  ``Possibilities of the X-ray diffraction data processing method for detecting
  reflections with intensity below the background noise component,''\\
  \emph{Powder Diffraction} \textbf{39}(3), 132--143 (2024).\\
  \href{https://doi.org/10.1017/S0885715624000241}{doi:10.1017/S0885715624000241}
  }
\end{center}

\vspace{0.75\baselineskip}

\noindent{\small
\textbf{Open Access License.}
This is an Open Access version of the article distributed under the terms of
the Creative Commons Attribution licence (CC BY 4.0), which permits unrestricted
re-use, distribution and reproduction in any medium, provided the original work
is properly cited.\\[2pt]
Copyright \textcopyright\ The Author(s), 2024.
First published in \emph{Powder Diffraction} by Cambridge University Press
on behalf of the International Centre for Diffraction Data.
}

\vspace{1.5\baselineskip}
%\setcounter{page}{1}%

\begin{abstract}
The values of the signal-to-noise ratio are determined, at which the method of processing X-ray diffraction data reveals reflections with intensity less than the noise component of the background. The possibilities of the method are demonstrated on weak reflections of $\alpha$-quartz. The method of processing X-ray diffraction data makes it possible to increase the possibilities of X-ray phase analysis in determining the qualitative phase composition of multiphase materials with a small (down to $0.1$ wt. \%) content of several (up to eight) phases.
\end{abstract}


\maketitle
\thispagestyle{empty}

\section{Introduction}

The method of X-ray phase analysis is widely used to study many types of materials (steels~\citep{Garin},  alloys~\citep{SarielDahan}, glass ceramics~\citep{Wilkins, Loy}, ceramics~\citep{Murugesan}, etc.) containing crystalline phases. The vast majority of researchers usually use a very high signal-to-noise ratio when registering reflections. The number of measured impulses is often in the range of hundreds of thousands for intense peaks and cannot be below tens of thousands for secondary peaks ~\citep{Guinebret}. However, in practice, the signal-to-noise ratio may not have large values, for example, $80-130$ ~\citep{Garin}, $\backsim$$70$ ~\citep{SarielDahan}, $30-60$ ~\citep{Wilkins}, $\backsim$$45$ ~\citep{Loy}.

The Symbol list below defines the symbols used in this manuscript.

-------------------------------------------------------------------

\textbf{Symbol list}

I is normalized intensity of reflection of the crystalline phase on the diffraction pattern;

${2\Theta}$ is the angle location of the crystalline phase reflection on the diffraction pattern;

d is the interplanar distance;

t is the counting time;

$\rm I_{b}$ is the intensity value at the ${2\Theta}$ angle of the diffraction pattern obtained without a specimen (diffractometer background in-tensity value);

$\rm I_{b-max}$ is the maximum background intensity;

$\rm I_{b-min}$ is the minimum background intensity;

 ($\rm I_{b-max} - I_{b-min}$) is the difference between the maximum and minimum values of the diffractometer back-ground intensities;

$\rm \langle I_{b}\rangle$ is the average diffractometer background intensity in the $[\Theta_{1} , \Theta_{2}]$ angle range;

$\rm \sigma_{b}$ is the rms deviation of the average value of the diffractometer background intensity in the angle range of $[\Theta_{1} , \Theta_{2}]$;

$\rm \Delta I_{b}$  is the ``noise'' component of the diffractometer background intensity in the $[\Theta_{1},\Theta_{2}]$ angle range;

$\rm I_{b}^{c}$  is the maximum intensity of pseudo-reflections on the correlation pattern after applying the data processing method to the intensity values of the diffractometer background in the $[\Theta_{1},\Theta_{2}]$ angle range;

${2\Theta}_{b}^{c}$ is the angle corresponding to the pseudo-reflection on the correlation pattern;

$\rm I_{q}^{d}$ is the $\alpha$-quartz reflection intensity on the diffraction pattern;

$\rm \beta$ is the FWHM;

$\rm I_{q}^{c}$  is the reflection intensity of $\alpha$-quartz on the correlation pattern after applying the data processing method;

${2\Theta}_{q}^{c}$  is the angle of the $\alpha$-quartz reflection on the correlation pattern after applying the data pro-cessing method;

$\rm \langle I_{q}^{d}\rangle$ is the average value of $\alpha$-quartz reflection intensity;

$\rm \langle I_{q}^{c}\rangle$ is the average intensity of $\alpha$-quartz reflection on the correlation pattern after applying the data processing method;

$\rm I_{q}^{d}$ / $
\rm \Delta I_{b}$ is the ratio of the $\alpha$-quartz reflection intensity to the value of the ``noise'' component of the diffractometer background intensity on the diffraction pattern (signal-to-noise ratio).

$\{{2 \rm \Theta_{i}; I_{i} }\}$,  $i=1$\ldots$ n, {n} \in {\mathbb{Z}}^{+}$ is the array of obtained experimental data (angles and intensities);

$\rm I^{(expr)} =\{ 2\Theta_{m+1}^{(expr)}$; $\ldots  \rm 2\Theta_{m+k}^{(expr)};\rm  I_{m+1}^{(expr)} \ldots \rm  I_{m+k}^{(expr)}\}$  is a limited sample from an array of experimental data in the range of angles from
$ \rm 2\Theta_{m+1}^{(expr)}$ to $ \rm 2\Theta_{m+k}^{(expr)}$;

$\rm I^{(mod)}=\{ 2\Theta_1^{(mod)}$\ldots$\rm 2\Theta_k^{(mod)};\rm  I_{1}^{(mod)}$\ldots$\rm  I_{k}^{(mod)}\}$ is a model data sampling in the range of angles from $\rm 2\Theta_1^{(mod)}$ to $\rm 2\Theta_k^{(mod)}$

$\rm S_{0}$ is the initial correlation degree between the model and experimental data samples;

$\rm I_{r}^{(rpm)}$ are intensity values from the experimental sample $\rm  I^{(expr)}$ after permutation of two intensity values;

$\pi$ is the permutation function;

$\rm S_{rpm j}$ is the current value of the correlation degree;

$\rm L_{j}$ is a numerical value of the difference between the current and initial values of the correlation degree $\rm L_{j}$=$\rm |S_{rpm j} -  S_{0}|$;

$\rm L_{max}$ is the maximum $\rm L_{j}$ value in the presence of a local maximum on the histogram;

$\rm I_{i}^{c}$ is an intensity on the correlation pattern in accordance with formula (\ref{eq:Iic});

$\{{2 \rm \Theta_{i}; I_{i}^{c} }\}$,  $i=1$\ldots$ n, {n} \in {\mathbb{Z}}^{+}$ is an array of correlation data obtained as a result of applying our method of processing X-ray diffraction data;

\textbf{Explanations on symbol indices.}
Symbols with index b refer to the background; q refers to $\alpha$-quartz  reflections; d refers to reflections on the diffraction pattern; c refers to reflections on the correlation pattern.

\textbf{Comments to the symbol list:}  $\rm I_{b}$, $\rm I_{q}^{d}$, $\rm 2\Theta_{q}^{d}$, are observed diffraction data; $\rm I_{b}^{c}$, $\rm 2\Theta_{b}^{c}$, $\rm I_{q}^{c}$, $\rm 2\Theta_{q}^{c}$ are calculated diffraction data:

----------------------------------------------------------------------------

However, situations are possible when researchers are forced to be satisfied with a low signal-to-noise ratio (values $3-4$ ~\citep{Loy}). This, for example, occurs if the time of studying the sample is limited when determining the phase composition and the content of the detected phases of steel directly in the process of its heat treatment ~\citep{Wiessner}. In that study the recording time of one diffraction pattern is limited to $700$ s. The signal-to-noise ratio was in the range from 7 to 1.5.

The same situation is possible when researchers are dealing with a multiphase material and where the content of several phases is low (down to $0.1$ wt. \%). The recording time of one diffraction pattern has to be increased not only to obtain reliable data on the reflection with low intensity but also due to a significant number of these reflections (possibly $10-20$ reflections from $3-5$ crystalline phases). For example, the authors studied lava-like fuel-containing materials (black ceramics) obtained as a result of the accident at the $4$th Unit of the Chornobyl nuclear power plant ~\citep{Gabielkov}. It was possible to obtain data on $78$ reflections, $76$ of which were identified from noise using the herein described method. From these reflections nine phases were identified. The signal-to-noise ratio was in the range from $4$ to $0.3$.

The signal-to-noise ratio can be improved for part of the diffraction pattern angular ranges. We can use the concept of a variable-counting-time (VCT) strategy ~\citep{MadsenHill}. This strategy is based on a function that increases the counting time used at each step in the scan in a manner that is inversely proportional to the decline in reflection intensity. This concept optimizes the acquisition time of diffraction data. However, it is not effective in the case of a large number of reflections with low intensity, i.e. for multiphase materials with a low content of the studied phases. You have to either increase the time it takes to obtain a diffraction pattern to unrealistic intervals or work with reflections that have a low signal-to-noise ratio.

In the practice of X-ray phase analysis, in most cases we are dealing with materials that contain one, two, and much less often three main crystalline phases, with the possible presence of several more phases in an amount of not more than a few percent (most often tenths of a percent or less). Data on phase composition and content of crystalline phases in small quantities can improve the characteristics of materials, and their parameters of preparation and use. The presence of crystalline phases with a low content in the material may be due to the presence of unavoidable impurities and/or specially introduced additives. Unfortunately, there can be several such phases (two, three or more) and their influence on the characteristics of the material is different. Some of them are useful, for example, they improve the characteristics of the material and simplify or reduce the cost of technology. The presence of others worsens the characteristics of the material. The influence of some of them has not yet been established.

To obtain more complete information about the phase composition of the material, it is desirable to identify crystalline phases with a low content and evaluate them. To do this, you can increase the recording time of the diffraction pattern or use a sensitive detector with low noise on the diffractometer, which increases the time or cost of research equipment. It is much more efficient to apply data processing methods, for example, our X-ray diffraction data processing method for multiphase materials with a low content of phases ~\citep{Skorbun}.

The purpose of this work is to determine the possibilities of the X-ray diffraction data processing method for multi-phase materials with low phase contents to identify weak reflections of crystalline phases with an intensity less than the background noise component.

\section{EXPERIMENTAL}
\begin{description}
\item[A.] Research materials

The study used a specimen of  $\alpha$-quartz powder placed in a cuvette. A specimen of brown LFCM ceramics, typical for room $304/2$ and the steam distribution corridor of the $4$th block of the Chornobyl nuclear power plant, was placed under the cuvette. The specimen of  $\alpha$-quartz powder was chosen to determine the possibilities of our diffraction data processing method. $\alpha$-quartz is a stable phase of silicon oxide and has been well studied. The specimen of LFCM brown ceramics was used to increase the background values in the diffraction pattern.

\item[B.] X-ray diffraction measurement

Data on $\alpha$-quartz reflections were obtained on a modernized DRON-$4$ diffractometer, scheme $\Theta$-2$\Theta$, Cu K$\alpha$ radiation ($30$ kV, $20$ mA), monochromator - graphite single crystal. The step size was $0.05^{\circ}$.  The counting times per point was $65$, $195$, and $390$ s (``large'' counting time) and from $5$ to $60$ s in increments of $5$ s (``small'' counting time). A screen was put in place to reduce the effect of $\gamma$-radiation from the studied specimens on the useful signal recording system. Addition-al protective barriers were using to protect personnel from $\gamma$-radiation of LFCM.

\item[C.] X-ray diffraction data processing method

X-ray diffraction data was processed by a fundamentally new method of correlation analysis using computational statistics ~\citep{Skorbun}. This method was specifically created for processing X-ray diffraction data from low-content multiphase materials. The method uses the permutation test, the Monte Carlo method and other computer statistics methods ~\citep{Moore}.

We create an array of experimental data $\{{2 \rm \Theta_{i}; I_{i} }\}$ from the corresponding values of angles and intensities. Next, we take part of this array for the values  $\rm i=m$\ldots$\rm m+k; {m,k} \in {\mathbb{Z}}^{+}$. This part of the data array may contain one reflection with a FWHM equal to the experimental width ($\beta$)  of the reflection.

Subsequently, the correlation value between the limited sample is calculated: \\ $\rm I^{(expr)} = \{ \rm 2\Theta_{m+1}^{(expr)};$\ldots$ \rm 2\Theta_{m+k}^{(expr)};\rm I_{m+1}^{(expr)}$\ldots$\rm I_{m+k}^{(expr)}\}$ from an array of experimental data with a model data sample $\rm I^{(mod)}=\{ \rm 2\Theta_1^{(mod)}$\ldots$\rm 2\Theta_k^{(mod)};\rm I_{1}^{(mod)}$\ldots$\rm  I_{k}^{(mod)}\}$, which models one reflection. The model sample $\rm I^{(mod)}$ is formed in the angle range from $\rm 2\Theta_1^{(mod)}$ to $\rm 2\Theta_{k}^{(mod)}$. A limited sample $\rm I^{(expr)}$ is taken from the same angle range, therefore: $\{\rm 2\Theta_{m+1}^{(expr)}$ \ldots $\rm 2\Theta_{m+k}^{(expr)}\} \equiv \{ \rm 2\Theta_1^{(mod)}$\ldots$\rm 2\Theta_{k}^{(mod)}\}$. We assign the value of the Gaussian function $\rm I^{(mod)}$ to each discrete reflection intensity value from the model sample:

\begin{equation}\label{eq:Ikm}
{\mathop{\rm I}\nolimits} _k^{(\bmod )}  =\rm I_0 e^{ - {\textstyle{{(2\Theta _{m + k}^{(\bmod )}  - 2\Theta _{m + k_0 }^{(\bmod )}) } \over {2\sigma ^2 }}}},
\end{equation}

where $\sigma^{2}$  is the dispersion; $k_{0}$ is the index at which the reflection angle 2${\Theta}$ corresponds to the maximum number of counts in a given range.

The sum of pairwise products of intensity values from the model $\rm I^{(mod)}$    and from the experimental  $\rm I^{(expr)}$ samples at the corresponding angle values gives a number characterizing the degree of correlation between the model and experimental data. The resulting number $\rm S_{0}$  will be called the initial value of the degree of correlation. According to the definition, this value has the form of a sum:
\begin{equation}\label{eq:So}
{\rm S_{0}= \sum_{l=1}^{l=k}\rm  I_{l}^{(mod)}  I_{m+l}^{(prm)}} .
\end{equation}
If two intensity values in the experimental sample are randomly permutated, we obtain
% формула 3

\begin{equation}\label{eq:Irprm}
$${\rm I}_{\rm r}^{\rm (prm)}  = \pi({\rm I}_{\rm r}^{\rm (expr)})  = \begin{array}{c}
   {\{ {\rm I}_{\rm m}^{\rm (expr)}  \ldots {\rm I}_{\rm p(i)}^{\rm (expr)}  \ldots {\rm I}_{\rm p(j)}^{\rm (expr)}  \ldots {\rm I}_{\rm m + k}^{\rm (expr)} \}  \to }  \\
  \! \!\!\!\!\!\!\!\! \{ {\rm I}_{\rm r}^{\rm prm}\,  \ldots {\rm I}_{\rm p(j)}^{\rm (expr)}  \ldots {\rm I}_{\rm p(i)}^{\rm (expr)}  \ldots {\rm I}_{\rm m + k}^{\rm (expr)} \} $$, \end{array}
\end{equation}

where $\pi$ is the permutation function, p(i) and p(j) are the indices of randomly selected array elements.

The new sum of pairwise products of intensity values from the model and the experimental sample changed in accordance with formula (\ref{eq:Irprm}) is $\rm S_(prm j)$= $\sum_{l=1}^{l=k}\rm I_{l}^{(mod)}  I_{m+l}^{(prm)}$. The resulting number $\rm S_(prm j)$ will be called the current value of the degree of correlation.

The numerical value of the difference between the current degree of correlation and the initial value will be called the $\rm L_{j}$ parameter. In accordance with the definition, this parameter takes the form:  $ \rm L_{j}$= $\rm|S_{prm  j}- S_{0} |$. A larger value of the parameter $\rm L_{j}$  corresponds to a larger correlation value between two samples $\rm I^{(expr)}$ and $\rm I^{(mod)}$ ). Let us carry out a large number (more than $10^5$ values) of permutations using a pseudorandom number generator in the experimental sample. For each current value of the degree of correlation, we obtain the corresponding values of the $\rm L_{j}$  parameter. The obtained current values of the $\rm L_{j}$  parameter are accumulated on a histogram. If a maximum is formed on the histogram, then the value of the $\rm L_{max}$ parameter is assigned to the intensity value on the correlation pattern. If there is no maximum in the histogram, the value $0$ is assigned to the intensity value. Thus, the intensity values on the correlation pattern are:
%формула 4
\begin{equation}\label{eq:Iic}
\rm I_{i}^{c}=\begin{cases} \rm  L_{max}, \; if  \; \exists \; L_{max} \\
{\quad} \ \, \, 0, \;  \rm  if  \; \nexists \;  L_{max}
               \end{cases}
\end{equation}

This processing algorithm is sequentially applied to each experimentally obtained intensity value $\rm I_{i}$ at the corresponding angle values ${\rm 2\Theta}_{i}$. Thus, a correlation pattern is formed based on the \{${\rm 2\Theta}_{i};\rm I_{i}^{c}\}$  array.

Subsequently, the correlation pattern is used to identify phases using well-known software, for example, Match! current version is $3.16$, developed by Crystal Impact (Germany) ~\citep{Putz}. The Crystallography Open Database (COD) diffraction database was used ~\citep{Duee}.

 \item[D.] Selection of weak $\alpha$-quartz reflections and angle ranges for their recording

           From all the reflections of $\alpha$-quartz, we chose $5$ reflections with a low relative intensity. The values of interplanar distances (d), angles (${2\Theta}$), and normalized intensities of these reflections ({I}) are given in Table~\ref{tab1}. These are both data obtained by us earlier ~\citep{Skorbun}  and data from the COD database ~\citep{Machatschki,Kroll}. The values of the intensities (normalized to $1000$) of the selected reflections are in the range $4.1-28.1$ rel. units. For subsequent experiments, the following intervals of ${2\Theta}$ angles were chosen: $55.2-56.2$ deg, $63.8-65.0$ deg, $73.2-74.2$ deg and $79.5-82.0$ deg, taking into account the positions of the selected reflections on the diffraction pattern.
% Table 1
\begin{table}[ht!]
\caption{Interplanar distance d, angles ${2\Theta}$, and normalized intensities I of low-intensity $\alpha$-quartz reflections}
	\centering
	\begin{tabular} { l c c c c c c c c c }
	
\hline
\hline

 \multicolumn{1}{l}{\strut  No.}    & \multicolumn{3}{c}{Authors experiment ~\citep{Skorbun}} & \multicolumn{3}{c}{COD 96-101-1777 ~\citep{Machatschki}} & \multicolumn{3}{c}{COD 96-153-2513 ~\citep{Kroll}} \\

\hline

\strut  \,  & d,~\AA & $2 \Theta$, & I,~rel. & d,~\AA & $2 \Theta$,deg & I,~rel. & d,~\AA & $2 \Theta$, & I,~rel. \\

\,  & \, & deg & units & \,  & \, & units & \,  &deg & units \\

\hline	

\strut 1 & 2 & 3 & 4 & 5 & 6 & 7 & 8 & 9 & 10 \\

 \hline

\strut 1. & 1.6530 &  55.60 & \, 5.2 & 1.6571 & 55.45 & 15.0 & 1.6537 & 55.58 & 16.7 \\

2. & 1.4482 & 64.30 & \, 7.4 & 1.4506 & 64.21 & 15.2 & 1.4477 & 64.65 & 17.6 \\

3. & 1.2855 & 73.70 & \, 7.4 & 1.2865 & 73.63 & 17.4 & 1.2838 & 73.81 & 22.2 \\

4.  & 1.1975 &  80.15 & 10.5 & 1.1975 & 80.15 & 25.1 & 1.1952 & 80.34 & 28.1 \\

5. & 1.1950 & 80.35 & \, 4.1 & 1.1946 & 80.38 & \, 8.1 & 1.1926 & 80.55 & \, 8.3 \\
 
\hline
\hline
\end{tabular}
\label{tab1}
\end{table}

\end{description}

\section{RESULTS }
\begin{description}

\item[A.]Characterization of the diffractometer background

First, we measured the background intensity of the diffractometer. To do this, we conducted measurements on the diffractometer without a sample in the selected angle intervals. Then, we processed the obtained data using our data processing method ~\citep{Skorbun}. Characteristics of the diffractometer background $\rm I_{b-max}$, $\rm I_{b-min}$, $\rm (I_{b-max}- I_{b-min})$, $\rm \langle I_{b}\rangle$, $\rm \sigma_{b}$ and $\rm \Delta {I}_{b}$ on the diffraction pattern and $\rm I_{b}^{c}$, $\rm 2\Theta_{b}^{c}$ on the correlation pattern are shown in Table~\ref{tab2}. A segment of the diffraction pattern is shown in  Figure~\ref{f1ab}({a}). A segment of the correlation pattern is shown in Figure~\ref{f1ab}(b) (results of diffraction data processing by the data processing method).
%TABLE II
\renewcommand{\arraystretch}{0.8}
		\begin{center}
\begin{table}[hb!]
\caption{Characteristics of the diffractometer background $\rm I_{b-max}$, $\rm I_{b-min}$, $(\rm I_{b-max}-I_{b-min})$, $\rm \langle I_{b}\rangle$, $\sigma_{b}$, ${\rm \Delta {I}_{b}}$ and background pseudo-reflections $\rm I_{b}^{c}$, $\rm 2\Theta_{b}^{c}$ obtained by data processing method (counting time 195 s, diffractometer without a specimen).}
\centering
\begin{tabular} { l c c c c c c c c r}

\hline
\hline

\strut No. & Angle in- & $\rm I_{b-max}$, & $\rm I_{b-min}$, & $\rm (I_{b-max}- I_{b-min})$, & $\rm \langle I_{b}\rangle$, & $\rm \sigma_{b}$, & $\rm \Delta {I}_{b}$, & $\rm I_{b}^{c}$, & $\rm 2\Theta_{b}^{c}$,  \\

\, & \quad terval, deg & counts & counts &  \qquad counts \qquad & counts & counts & counts & counts & deg \\

\hline

\strut 1 & 2 & 3 & 4 & 5 & 6 & 7 & 8 & 9 & 10  \\

\hline

\strut 1. & 55.2--56.2 & 21.2 & 8.9 & 12.3 & 14.6 &3.7 & 14.8 &  1.6 & 55.6 \\

 2. & 63.8--65.0 & 22.7 & 8.0 & 14.7 & 14.9 & 3.5 &  14.0 & 3.5 & 64.7 \\

 3. & 73.2--74.2 & 25.5 & 8.9 & 16.6 & 16.6 & 4.0 & 16.0 & 3.6 & 73.5 \\

 4. & 79.5--82.0 & 25.2 & 5.6 & 19.6 & 14.2 & 4.6 & 18.4 & 4.0 & 81.3 \\
 
 \hline
 \hline

\end{tabular}
\label{tab2}
\end{table}
\end{center}
%Fig. 1. 
\begin{figure}[ht!]
  \centering
  \vspace{1mm}
  \includegraphics[width=0.8\textwidth]{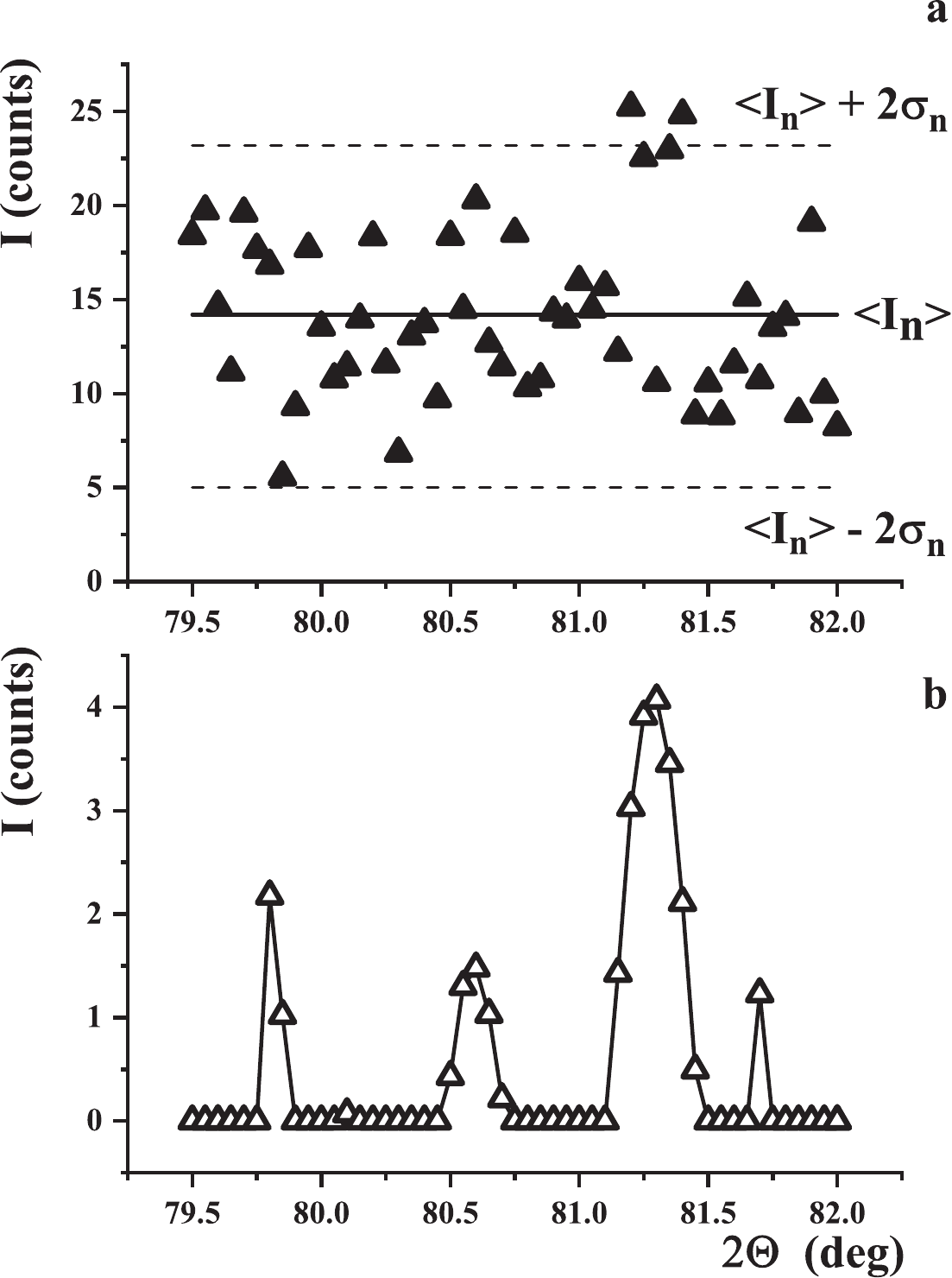}%
  \vspace{-1mm}
  \caption{Fragments of (a) diffraction and (b) correlation patterns of the diffractometer background (exposure time 195 s).}
  \label{f1ab}
\end{figure}

The dependence of the background intensity on the angle has the form of a wide band ( Figure~\ref{f1ab}(a)). It has a constant component and a ``noise'' component. For each angle intervals (Table~\ref{tab2}, column {2}), the maximum background intensity values $\rm I_{b-max}$ (column {3} are in the range of $21.2-25.5$ counts, the minimum $\rm I_{b-min}$ (constant component) - $5.6-8.9$ counts, their difference $\rm I_{b-max}-\rm I_{b-min}$, (noise component, column {5}) $12.3-19.6$ counts. With this approach, the background components depend on specific maximum and minimum values. We will try to use all background values within the same angle interval. Let us find the average values of the background intensities $\rm \langle I_b \rangle$ (Table~\ref{tab2}, column {6}) and standard deviations $\rm \sigma_{b}$ (column {7}). The amplitude of the ``noise'' component of the background $\rm \Delta I_{b}$ (column 8) is equal to the quadruple standard deviation of the background average intensity values in accordance with formula ~$\rm \Delta I_{b}= 4 \rm \sigma_{b}$. In the range of values from $\rm \langle I_{b}\rangle$- 2$\rm \sigma_{b}$ to $\rm \langle I_{b}\rangle$+2$\rm \sigma_{b}$, the background intensities are found with a probability of ~$95.4$ $\%$. This is clearly seen in Figure~\ref{f1ab}(a). Of the $40$ experimental points shown in the figure, $38$ points are located inside the band with a ``width'' of 4$\rm \sigma_{b}$, i.e. 95 $\%$. Now the characteristics of the background are practically independent of specific maximum and minimum values. They are determined considering the entire data array for a given angle interval.

A segment of the correlation pattern was obtained by applying our data processing method to the diffraction pattern (Figure~\ref{f1ab}{b}). The constant component of the background was ``cut off'', and the ``noise'' component was converted into several pseudo-reflections. The intensity values of pseudo-reflections $\rm I_{b}^{c}$ (Table~\ref{tab2}, column {9}) for all angle intervals do not exceed {4} counts.

We see (Table~\ref{tab2}) that the average values of the background intensity of the diffractometer ~$\rm \langle I_{b}\rangle$ (column 6) and the values of the ``noise'' component of the background $\rm \Delta I_{b}$ (column {8}) are in the range of $14.2 - 16.6$ counts and $14.0 - 18.4$ counts respectively for all angle intervals with the counting time of $195$ s. In this case, applying the method of processing diffraction data to the intensities of the diffractometer background for a range of angles gives pseudo-reflections $\rm I_{b}^{c}$ (column $9$) with an intensity of up to $4$ counts.

\item[B.]Determination of signal-to-noise ratios for weak $\alpha$-quartz reflections

Next, we mounted a specimen of $\alpha$-quartz on a diffractometer. Measurements were carried out in the intervals of angles No. $1-4$ (see Table~\ref{tab3}, column $2$). Table~\ref{tab3} presents the obtained characteristics of $\alpha$-quartz  and background reflections: (intensities and angles): $\rm I_{q}^{d}$, $\rm 2\Theta_{q}^{d}$, $\rm \langle I_{b}\rangle$, $\rm \sigma_{b}$, $\rm \Delta I_{b}$, and $\rm I_{q}^{d}$/$\rm \Delta I_{b}$  signal-to-noise ratios. The resulting data was then processed using our data processing method. The obtained characteristics of $\alpha$-quartz reflections $\rm I_{q}^{c}$, $\rm 2\Theta_{q}^{c}$ and background pseudo-reflections $\rm I_{b}^{c}$ re also shown in Table~\ref{tab3}. A segment of the diffraction pattern with the reflection of $\alpha$-quartz is presented in Figure {2}(a), correlation pattern - in Figure~\ref{f2ab}(b). Nonlinear curve fit of each reflection of $\alpha$-quartz to a Gaussian function was carried out on the diffraction pattern (Figure~\ref{f2ab}a).
%Таблиця 3
\begin{table}[hb!]
\caption{Characteristics of $\alpha$-quartz reflections $\rm I_{q}^{d}$, $\rm 2\Theta_{q}^{d}$ and background $\rm \langle I_{b}\rangle$, $\rm \sigma_{b}$, ${\rm \Delta {I}_{b}}$, as well as  $\alpha$-quartz reflections $\rm I_{q}^{c}$ , $\rm 2\Theta_{q}^{c}$ and background pseudo-reflections $\rm I_{b}^{c}$ obtained by the method data processing; $\rm I_{q}^{d}$)/$\rm \Delta {I}_{b}$  ratios, (counting time 195 s, $\alpha$-quartz specimen on the diffractometer).}
\centering
 \footnotesize
\begin{tabular} { l c c c c c c c c c c } 
\\
\hline
\hline

\strut No. & Angle & $\rm  I_{q}^{d}$, & $\rm 2\Theta_{q}^{d}$, & $\rm  I_{q}^{c}$, & $\rm 2\Theta_{q}^{c}$, & $\rm  \langle I_{b}\rangle$, & $\rm \sigma_{b}$, & $\rm  \Delta {I}_{b}$, & $\rm I_{b}^{c}$, & $\rm  {I_{q}^{d}}/{\rm \Delta {I}_{b}}$   \\

\, &  interval, deg &counts & deg & counts & deg & counts & counts & counts & counts & \,  \\

\hline

\strut 1 & 2 & 3 & 4 & 5 & 6 & 7 & 8 & 9 & 10 & 11 \\
\hline

\strut 1. & 55.2--56.2 & $20.0\pm3.5$ & $55.46\pm0.02$ & 10.5 & $55.45\pm0.02$ & 19.7 & 2.7 &  10.8 & 1.5 & 1.85  \\

 2. & 63.8--65.0 & $14.9\pm2.4$ & $64.15\pm0.03$ & \, 7.5 & $64.15\pm0.02$ & 17.6 &  2.4 & \, 9.6 & 1.5 & 1.55 \\

 3. & 73.2--74.2 & $24.7\pm4.2$ & $73.62\pm0.02$ & 11.0 & $73.65\pm0.02$ & 17.8 & 3.8 & 15.2 & 3.7 & 1.63 \\

 4. & 79.5--82.0 & $20.0\pm3.5$ & $80.10\pm0.02$ & 11.7 & $80.10\pm0.02$ & 20.8 & 3.5 & 14.0 & 0.0 & 1.94 \\

 \, & \qquad & $35.9\pm8.8$ & $81.54\pm0.02$ & \, 9.4 & $81.60\pm0.02$ & 27.5 & 2.2 & 8.8 & 0.0 & 4.10 \\

\hline
\hline 
\end{tabular}
\label{tab3}
\begin{flushleft}
	  \footnotesize{Note: the half-width of reflections ($\beta$) was 0.3 degrees.} \\
    \end{flushleft}
\end{table}
%Figure 2. 
\begin{figure}[ht!]
  \centering
  \vspace{1mm}
  \includegraphics[width=0.7\textwidth]{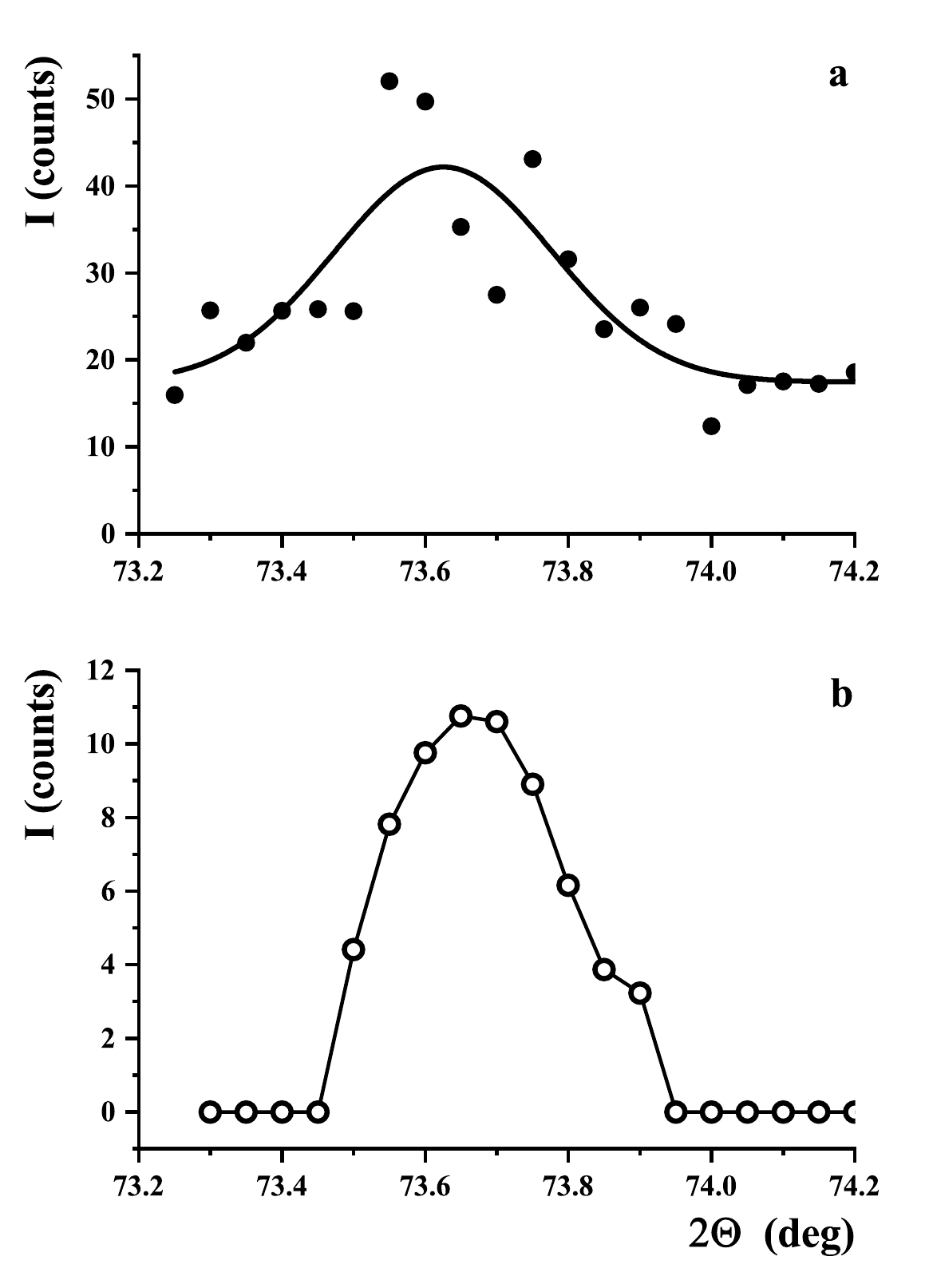}%
  \vspace{-1mm}
  \caption{Fragments of the diffraction pattern (Fig.~\ref{f2ab}(a)) and correlation pattern (Fig.~\ref{f2ab}(b)) of $\alpha$-quartz reflection (exposure time 195 s).}
  \label{f2ab}
\end{figure}

The values of the intensities of the $\alpha$-quartz reflections $\rm I_{q}^{d}$, determined by the nonlinear curve fit to a Gaussian function, are in the range of $14.9-35.9$ counts (Table~\ref{tab3}, column {3}). The values of the $\rm 2\Theta_{q}^{d}$ angles are presented in column $4$. The values of the intensities of $\alpha$-quartz reflections after applying our method of data processing on the correlation pattern $\rm I_{q}^{c}$ (column $5$) are $7.5-11.7$ counts, which is $\rm \sim2$ times less than the intensity of $\alpha$-quartz reflections $\rm I_{q}^{d}$ (column $3$). The angles $\rm 2\Theta_{q}^{c}$ at which the reflections are located on the correlation pattern (column $6$), correspond to the values on the diffraction pattern (column $4$) taking into account the determination error. The average background intensities $\rm \langle I_{b}\rangle$ increased $1.5$ times to $17.3-27.5$ counts (column $7$) compared with the average values of the background intensities of the diffractometer (Table~\ref{tab2}, column $6$). This is due to the fact that scattering from the  $\alpha$-quartz specimen also contributes to the background intensity. Standard deviations of the background intensity values $\rm \sigma_{b}$ (column $8$) decreased to $2.2-3.8$ counts, and the values of the background ``noise'' component $\rm \Delta I_{b}$ (column $9$) decreased to $8.8-15.2$ counts. The intensity values of pseudoreflections after applying the data processing method ($\rm I_{b}^{c}$) (column $10$) do not exceed $4$ counts. After applying the data processing method, the reflection intensities of  $\alpha$-quartz $\rm I_{q}^{c}$ (column $5$) significantly exceed the intensities of pseudoreflection $\rm I_{b}^{c}$ (column $10$), which indicates their high reliability. The ratios of the reflection intensities of  $\alpha$-quartz $\rm I_{q}^{d}$ (column $3$) to the values of the ''noise'' component of the background $\rm \Delta I_{b}$ (column $9$) are in the range of $1.55-4.10$ (column $11$), i.e. the intensity of reflections of  $\alpha$-quartz is significantly greater than the value of the ``noise'' component of the background $\rm \Delta I_{b}$. Visually, this is manifested in the fact that the reflections of  $\alpha$-quartz stand out above the background (Figure~\ref{f2ab}(a)).

The signal-to-noise ratios for weak $\alpha$-quartz reflections are significantly greater than $1$. In this case, the $\alpha$-quartz reflections stand out above the background. The data processing method relia-bly identifies $\alpha$-quartz reflections. The intensity of $\alpha$-quartz reflections is significantly greater than the intensity of background pseudo-reflections.

\item[C.]Increasing background characteristics due to an additional source of $\gamma$-quanta from brown LFCM ceramics

A specimen of LFCM brown ceramics was placed near the cuvette with a specimen of $\alpha$-quartz. This was done to increase the background due to $\gamma$-quanta, which are emitted by LFCM radionuclides. Then, the intervals of angles were chosen, in which there are no reflections of $\alpha$-quartz. This made it possible to measure the background in the presence of the LFCM specimen. The selected angle intervals are shown in Table~\ref{tab4} (column $2$). It can be seen that for a counting time of $195$ s, all indicators such as: average values of the background intensity $\rm \langle I_{b}\rangle$ (column $4$), standard deviations of the values of the background intensity $\rm \sigma_{b}$ (column $5$), values of the ``noise'' component of the background ($\rm \Delta I_{b}$, column $6$) - increased compared to the values of the corresponding parameters of the background of the diffractometer without specimens of $\alpha$-quartz and brown ceramics (see Table~\ref{tab2}). The average value of the background $\rm \langle I_{b}\rangle$ (column $4$) increased significantly by $1.7-1.9$ times, and the values of the ``noise'' component of the background $\rm \Delta I_{b}$ (column $6$) did not increase significantly - by $1.05-1.17$ times. After applying our data processing method, the value of the intensity of pseudo-reflections $\rm I_{b}^{c}$ (column $7$) also does not exceed values of more than $4$ counts for a counting time of $195$ s (except one value in the range of angles $7.0-10.0$). The positions of pseudo-reflections after applying the data processing method (the corresponding angles ($ \rm 2\Theta_{b}^{c}$) for one interval of angles do not match as they should be (see Table~\ref{tab4} , column $8$). The correlation pattern with background pseudo-reflections is shown in Fig. $3$ b.
%Таблиця 4
\begin{center}
\begin{table}[ht]
\caption{Characteristics of the diffractometer background $\rm \langle I_{b}\rangle$, $\rm \sigma_{b}$, ${\rm \Delta {I}_{b}}$ and background pseudo-reflections $\rm I_{b}^{c}$, $\rm {2\Theta}_{b}^{c}$ obtained by data processing (counting time (t) 65 and 195 s; specimens of $\alpha$-quartz and LFCM brown ceramics are on the diffractometer).}
\centering
\begin{tabular} { l c c c c c c c }

\hline
\hline

 $\rm \strut No.$ & $\rm Angle$ & $\rm  t$, &  $\rm  \langle I_{b}\rangle$, & $\rm \sigma_{b}$, & $\rm  \Delta {I}_{b}$, & $\rm I_{b}^{c}$, &  $\rm 2\Theta_{b}^{c}$, \\
 
\, & \, interval, deg & s & counts & counts & counts & counts & deg \\
 
\hline

\strut 1 & 2 & 3 & 4 &5 &  6 & 7 & 8   \\

\hline

\strut 1. & \, 7.0--10.2  & 65  &  13.3 &  2.9  &  11.6  &  1.2  &  \, 9.35 \\

\, & \, & 195 & 39.9 & 4.8 & 19.2 & 4.5 & \, 8.20  \\

 2. & 61.0--63.0 & 65 & \, 9.2 & 1.9 & \, 7.6 & 2.0 &  61.95   \\

  &  & 195 & 28.2 & 3.7 & 14.7 & 2.7 & 61.90   \\

 3. & 69.0--71.0 & 65 & \, 9.4 & 2.6 & 10.4 & 1.4 & 69.75  \\

  &  & 195 & 28.7 & 4.6 & 18.4 & 2.5 & 61.30   \\

 4. & 78.0--80.0 & 65 & 11.1 & 3.9 & 15.6 & 0.6 & 79.25  \\

  &  & 195 & 29.4 & 6.1 & 24.4 & 2.1 & 78.85   \\

\hline
\hline 
\end{tabular}
\label{tab4}
\end{table}
\end{center}
% Figure 3.
\begin{figure}[ht!]
  \centering
  \vspace{1mm}
  \includegraphics[width=0.7\textwidth]{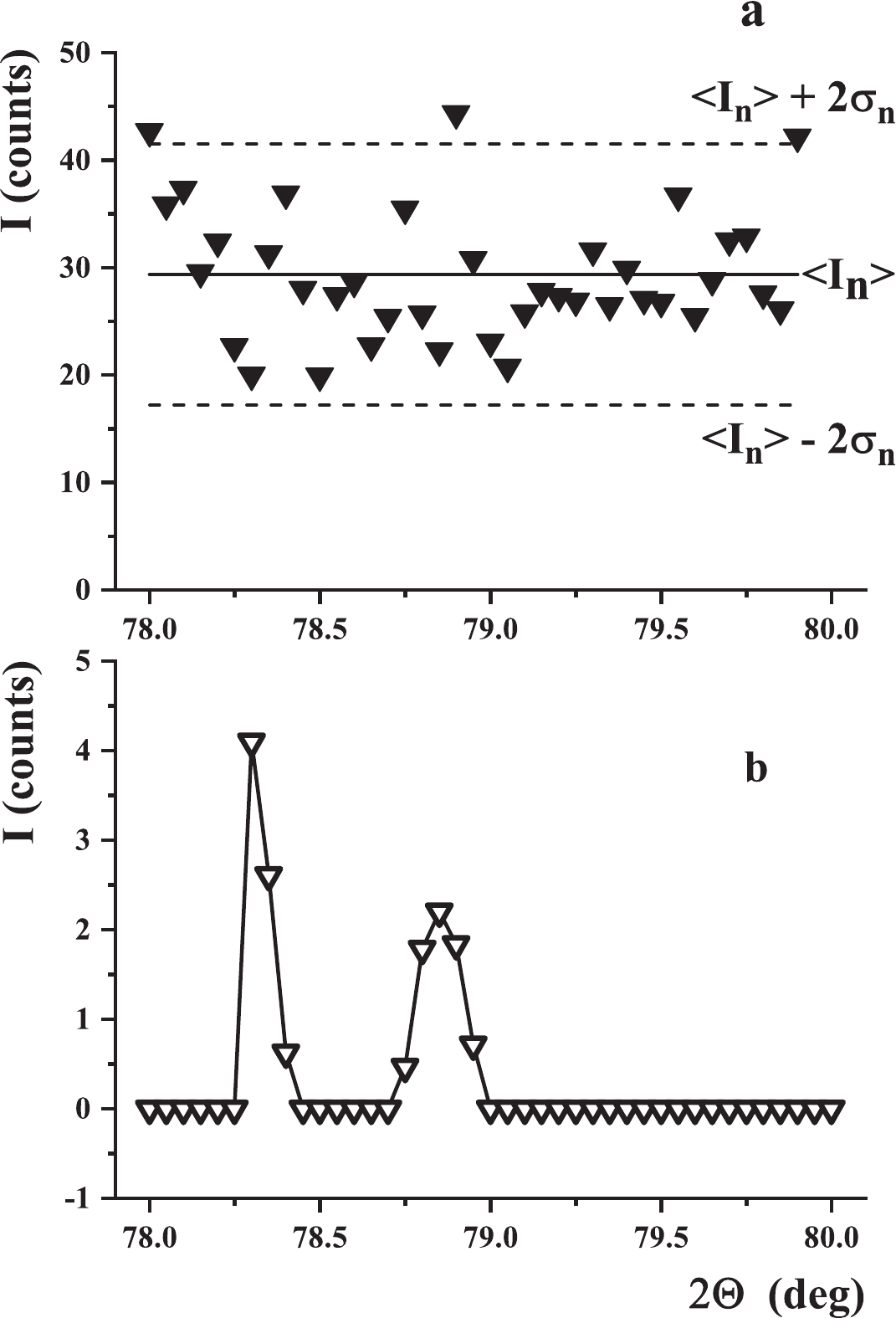}%
  \vspace{-1mm}
  \caption{Fragments of the diffraction pattern (Fig.~\ref{f3ab}(a)) and correlation pattern (Fig.~\ref{f3ab}(b)) of the background from the diffractometer with specimens of $\alpha$-quartz and brown ceramics (exposure 195 s).}
  \label{f3ab}
\end{figure}

Thus, the average value of the background intensity and the value of the ''noise'' component of the background increased due to the use of an additional source of $\gamma$-quanta from brown LFCM ceramics.

\item[D.]Determination of signal-to-noise ratios for weak $\alpha$-quartz reflections with enhanced back-ground

Now we have returned to the same angle intervals (see Table~\ref{tab3}, column $2$), which contain reflections of $\alpha$-quartz. The measurements were carried out at three counting times (Table~\ref{tab5}, column $3$). We have obtained reflections of . One of the diffraction patterns is shown in Figure~\ref{f4ab}(a), and a correlation pattern in Figure~\ref{f4ab}(b). We see that, as it should be, for each angle interval with an increase in the recording time from $65$ s to $195$ s and up to $390$ s, the reflection intensity increases by $\sim$ $3$ and $\sim$ $6$ times (column $4$). The values of the angles $\rm 2\Theta_{q}^{d}$ corresponding to them practically coincide, taking into account the data processing error (column $5$). After applying the data processing method $\rm I_{q}^{c}$ (column $6$), the reflection intensities are $\sim$ $2$ times less than the reflection intensities of $\alpha$-quartz $\rm I_{q}^{d}$ for each of the three counting times in all angle intervals (column $4$). The values of the angles corresponding to them $\rm2\Theta_{q}^{c}$ (column $7$) practically coincide with the angles for reflections of $\alpha$-quartz $\rm 2\Theta_{q}^{d}$ (column $5$). The average values of the background intensities  $\rm \langle I_{b}\rangle$ (column $8$) increase by $\sim$ $3$ and $\sim$ $6$ times for each angle interval with increasing recording time from $65$ s to $195$ s and up to $390$ s. The values of the ``noise'' component of the background $\rm \Delta I_{b}$ (column $10$) also increase, but to a lesser extent. The maximum intensity of pseudoreflections after applying the data processing method $\rm I_{b}^{c}$ (column $11$) does not exceed $2.4$ counts. After using the data processing method, the maximum intensity of background reflections $\rm I_{b}^{c}$ (column $11$) does not exceed $2.4$ counts for a counting time of $65$ s; $5.7$ counts (greater than $4$, outlier) for a counting time of $195$ s and $8.0$ counts for a counting time of $390$ s, respectively. The ratio of the reflection intensity of $\alpha$-quartz to the values of the ``noise'' component of the background $\rm I_{q}^{d}$ / $\rm \Delta I_{b}$  (column $13$) ranges from $0.56$ to $1.94$. By changing the counting time, we can change the ratio of the reflection intensity values of $\alpha$-quartz to the values of the ''noise'' component of the background $\rm I_{q}^{d}$ / $\rm \Delta I_{b}$.

%Таблиця 5
\begin{table}[ht!]
\caption{Characteristics of $\alpha$-quartz reflections $\rm I_{q}^{d}$, $\rm 2\Theta_{q}^{d}$   and background $\rm \langle I_{b}\rangle$, $\rm \sigma_{b}$, ${\rm \Delta {I}_{b}}$; $\alpha$-quartz reflections $\rm I_{q}^{c}$ , $\rm 2\Theta_{q}^{c}$  and background pseudo-reflections $\rm I_{b}^{c}$ obtained by data processing; ratios $\rm I_{q}^{d}$/${\rm \Delta {I}_{b}}$ (counting time (t) 65, 195 and 390 s, specimens of $\alpha$-quartz and brown LFCM ceramics on the diffractometer).}
\centering
 \footnotesize
\begin{tabular} { l c c c c c c c c c c c c }

\hline
\hline

\strut No. & Angle & $\rm t,$ & $\rm I_{q}^{d}$, & $\rm 2\Theta_{q}^{d}$, & $\rm I_{q}^{c}$, & $\rm 2\Theta_{q}^{c}$, & $\rm \langle I_{b}\rangle$, & $\rm \sigma_{b}$, & $\rm \Delta {I}_{b}$, & $\rm I_{b}^{c}$ & $\rm 2\Theta_{b}^{c}$, & $\rm I_{q}^{d}$/${\rm \Delta {I}_{b}}$, \\

\, \strut & interval, & s & counts & deg & counts & deg & counts & counts & counts & counts &  deg & \, \\

\, & deg & \, & \, & \, & \, & \, & \, & \, & \, & \, & \, & \, \\

\hline

\strut 1 & 2 & 3 & 4 & 5 & 6 & 7 & 8 & 9 & 10 & 11 & 12 & 13 \\ 

\hline

\strut 1. & 55.2--56.2  & 65 &  \, $5.9 \pm \, 1.8$ &  $55.41\pm0.02$ & 2.7 &  55.40  & 10.5 & 2.5 & 10.0 & 1.0 & 55.80 & 0.59 \\

 &  & 195 &  $17.9 \pm \, 4.6$ &  $55.40\pm0.02$ & 8.5 &  55.40  & 28.4 & 4.7 & 18.8 & 2.5 & 55.80 & 0.95 \\

 &  & 390 &  $38.1 \pm \, 6.4$ &  $55.42\pm0.01$ & \!\!\!18.0 &  55.40  & 57.6 & 6.3 & 25.2 & 7.5 & 55.80 & 1.51 \\

2. & 63.8--65.0  & 65 &  \, $6.4 \pm \,1.4$ &  $64.24\pm0.03$ & 3.3 &  64.15  & 10.5 & 2.2 & 8.8 & 2.4 & 64.50 & 0.73 \\

 &  & 195 &  $16.9 \pm \, 3.7$ &  $64.17\pm0.04$ & 7.7 &  64.15  & 25.4 & 4.3 & 17.2 & 3.5 & 64.55 & 0.98 \\

 &  & 390 &  $32.8 \pm \, 5.5$ &  $64.19\pm0.03$ & \!\!\!17.5 &  64.15  & 52.6 & 5.3 & 21.2 & 8.0 & 64.50 & 1.55 \\

3. & 73.2--74.2  & 65 &  \, $6.1 \pm \, 2.0$ &  $73.70\pm0.05$ & 2.0 &  73.65  & 11.5 & 2.7 & 10.8 & 2.2 & 73.90 & 0.56 \\

 &  & 195 &  $17.9 \pm \, 3.5$ &  $73.64\pm0.03$ & 8.0 &  76.60  & 28.6 & 3.8 & 15.2 & \,\,\, 5.7$^{\ast \ast}$ & 73.90 & 1.18 \\

4. & 79.5--82.0  & 65 &  $11.6 \pm \, 2.2$ &  $80.11\pm0.03$ & 5.5 & 80.15 & \, 9.6 & 2.8 & 11.2 & - & - & 1.04 \\

 &  & 195 &  $29.2 \pm \, 4.8$ & $80.10\pm0.03$ & \!\!\!13.0 &  80.10  & 29.5 & 5.2 & 20.8 & - & - & 1.04 \\

 &  & 390 &  $55.3 \pm \, 9.3$ & $80.09\pm0.03$ & \!\!\!23.0 &  80.15  & 58.6 & 8.2 & 32.8 & - & - & 1.40 \\

 &  & 65 &  -$^{\ast}$ & - & 3.0 &  81.60  & \, 9.6 & 2.8 & 11.2 & - & - & - \\

 &  & 195 &  $32.2 \pm \, 9.3$ & $81.55\pm0.02$ & \!\!\!12.5 &  81.60  & 29.6 & 5.1 & 20.4 & - & - & 1.58 \\

 &  & 390 & $63.7\pm16.8$ & $81.55\pm0.01$ & \!\!\!25.0 &  81.60  & 58.6 & 8.2 & 32.8 & - & - & 1.94 \\

\hline
\hline 

\end{tabular}
\label{tab5}
\begin{flushleft}
	  \footnotesize{Note: the half-width of reflections ($\beta$)is equal to 0.3 deg; \\
	  
	  $^{\ast}$ - experimental data did not fit to a Gaussian function; \\
	  
	  $^{\ast \ast}$ - spike.  } \\
    \end{flushleft}
\end{table}

Average background intensities $\rm \langle I_{b}\rangle$ for the $\alpha$-quartz and the brown LFCM ceramic specimens (Table~\ref{tab5}, column $8$, counting time $195$ s) compared to the average background intensities for the $\alpha$-quartz specimen without the brown LFCM ceramic specimen (Table~\ref{tab3}, column $7$) increased by an average of $\sim$ $1.4$ times due to the use of the brown LFCM ceramic specimen as a source of ``noise'' gamma quanta.

The values of the ``noise'' component of the background $\rm \Delta I_{b}$ for a specimen of $\alpha$-quartz with a specimen of brown ceramics LFCM (Table~\ref{tab5}, column $10$, counting time $195$ s) compared with the values of the ``noise'' component of the background for the specimen of $\alpha$-quartz without the specimen of brown ceramics LFCM (Table~\ref{tab3}, column $9$) increased by an average of $\sim$ $1.7$ times due to the use of the specimen of brown ceramics.

We see (Figure~\ref{f4ab}(a)) that the reflection of $\alpha$-quartz in the range of angles $63.8-64.8$ deg at a counting time of $195$ s, stands out well against the noise component of the background (the ratio of the intensity of $\alpha$-quartz reflection to the values of the ''noise'' component of the background $\rm I_{q}^{d}$ / $\rm \Delta I_{b}$ is $0.98$). After applying our method of processing diffraction data (Figure~\ref{f4ab}(b)), this reflection is recorded with high reliability in the correlation pattern (its intensity is greater than $4$ counts).
%Figure 4. 
\begin{figure}[ht!]
  \centering
  \vspace{1mm}
  \includegraphics[width=0.7\textwidth]{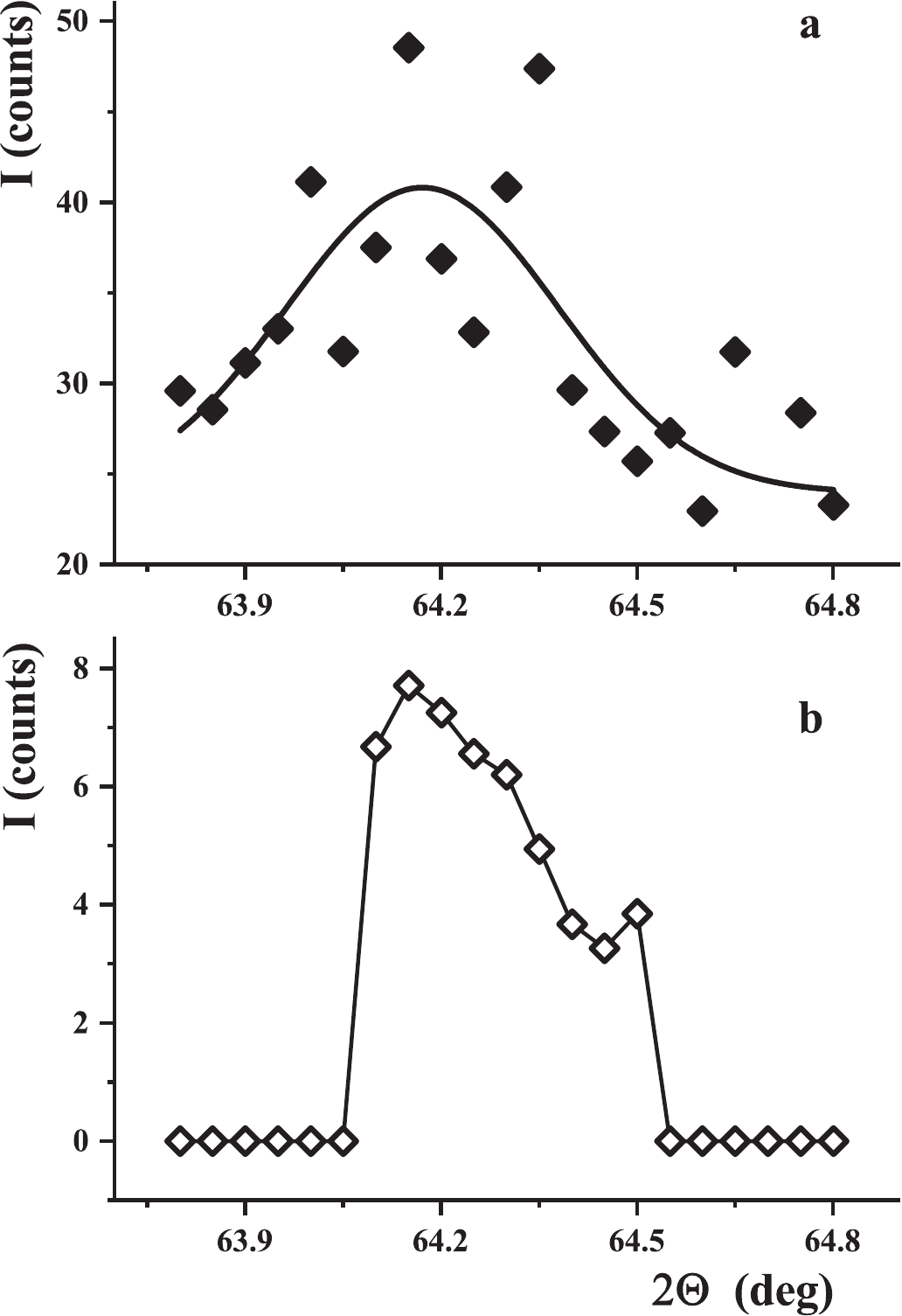}%
  \vspace{-1mm}
  \caption{Fragments of the diffraction pattern (Fig.~\ref{f4ab}(a)) and correlation pattern (Fig.~\ref{f4ab}(b)) of $\alpha$-quartz with brown ceramics (exposure 195 s).}
  \label{f4ab}
\end{figure}

The signal-to-noise ratios for weak $\alpha$-quartz reflections were reduced to $0.95-1.58$ at a count-ing time of $195$ s compared to the previously obtained values of $1.55-4.1$ in section $5.3$. This was possible due to an increase in the intensity of the ``noise'' component of the background us-ing an additional source of $\gamma$-quanta from brown LFCM ceramics.

The best results in this section are signal-to-noise ratios of $0.56-0.73$ (significantly less than $1$). These were obtained with a counting time of $65$ s (see Table~\ref{tab5}, angle intervals No. $1$, $2$ and $3$).

\item[E.]Identifying the $\alpha$-quartz reflection with a signal-to-noise ratio less than $1$. Refinement of characteristics

Next, let us pay attention to the cases when the intensity of reflections of $\alpha$-quartz $\rm I_{q}^{d}$ (Table~\ref{tab5}, column $4$) is less than the ``noise'' component of the background $\rm \Delta I_{b}$ (Table~\ref{tab5}, column $10$). These are the angle intervals $55.2-56.2$, $63.8-65.0$ and $73.2-74.2$ deg, counting time $65$ s. In this case, the ratio of the intensity of $\alpha$-quartz reflection to the values of the ``noise'' component of the background $\rm I_{q}^{d}$ / $\rm \Delta I_{b}$  is in the range of $0.56-0.73$. Let's choose one of the cases-the interval of angles $63.8-65.0$ deg with a counting time of $65$ s.

We took $6$ measurements in this range of angles with a counting time of $60$ s. The $\rm I_{q}^{d}$, $\rm 2\Theta_{q}^{d}$ characteristics of $\alpha$-quartz reflections, also $\rm I_{q}^{c}$ , $\rm 2\Theta_{q}^{c}$ after applying the data processing method, are presented in Table~\ref{tab6} (angle interval $63.8-64.8$ degrees, counting time $60$ s).
% Table 6
\begin{table}[ht!]
\caption{Characteristics of $\alpha$-quartz reflections $\rm I_{q}^{d}$, $\rm 2\Theta_{q}^{d}$, and $\rm I_{q}^{c}$ , $\rm 2\Theta_{q}^{c}$ obtained by data processing (in the angle interval $63.8-64.8$ deg with a counting time of 60 s, the specimens of $\alpha$-quartz and brown LFCM ceramics are on a diffractometer). }
\centering
\begin{tabular} { l c c c c }

\hline
\hline 

\strut No.  & \quad $\rm I_{q}^{d}$, & $\rm 2\Theta_{q}^{d}$, & $\rm I_{q}^{c}$, & $\rm 2\Theta_{q}^{c}$, \\

\strut \, & counts & deg &counts & deg  \\

\hline 

\strut 1 & \, 2 \,\,\,  & \,\, 3 \,\,  & 4 & 5  \\

\hline 

\strut 1. & 6.52  & 64.18 & 1.75 &  64.24 \\
  
2. & 7.29  & 64.15 & 3.04 & 64.25 \\

3. & 6.35 & 64.24 &2.73 &  64.24 \\

4. & 6.51 & 64.15 & 2.79 & 64.27 \\

5. & 5.88 & 64.17 & 2.82 & 64.21 \\
 
6. & 8.07 & 64.15 & 5.04 & 64.14 \\

$\langle { \,\,\,}\rangle$ & 6.77 & 64.17 & 3.03 & 64.22  \\

\,$\sigma$ & \,0.80 &\, 0.04 & 1.08 &  0.05 \\	
	
\hline
\hline

\end{tabular}
\label{tab6}
\begin{flushleft}
\centering
	  \footnotesize{Note: $\langle { \,\,\,}\rangle$ denotes average values of  $\rm I_{q}^{d}$, $\rm 2\Theta_{q}^{d}$, $\rm I_{q}^{c}$, $\rm 2\Theta_{q}^{c}$.} \\
	     \end{flushleft}
\end{table}

We see (Table~\ref{tab6}, column $2$) that the intensity of $\alpha$-quartz reflections $\rm I_{q}^{d}$ is $5.88-8.07$ counts. The average $\rm \langle I_{q}^{d}\rangle$   value is  $6.77\pm0.8$ counts. The angles corresponding to these reflections are in the range of $64.15-64.27$ degrees. The average value is $64.17\pm0.04$ deg. After applying our method of processing X-ray diffraction data, the reflection intensities $\rm I_{q}^{c}$ are in the range of $2.73-5.04$ counts. The average $\rm \langle I_{q}^{c}\rangle$ value is $3.03\pm1.08$ counts. The angles $\rm 2\Theta_{q}^{c}$ corresponding to these reflections are in the same range of $64.14-64.27$ deg. The average value is $64.22\pm0.05$ deg. After applying the method, the reflection intensity decreased by $2.2$ times, and the average values of the reflection angles did not change (considering the measurement error).

For subsequent use, we chose the $\alpha$-quartz reflection in the angle range of $63.8-64.8$ degrees with a counting time of $60$ s. The refined reflection intensity is $6.77\pm0.8$ counts.

\item[F.] Determination of the signal-to-noise ratio values at which the data processing method allows the $\alpha$-quartz reflection to be separated from the background noise component

To elucidate the possibilities of our method of processing X-ray diffraction data, we recorded the reflections of $\alpha$-quartz in the angle range of $63.8-64.8$ deg at different counting times (reducing the counting time from $60$ to $5$ s), thereby providing the ratio of the intensity of $\alpha$-quartz reflection to the values of the ``noise'' component of the background $\rm I_{q}^{d}$ / $\rm \Delta I_{b}$ less than $0.73$ (Table~\ref{tab5}, column $13$). The characteristics of the background were determined from the recording results in the angle range $61.0-63.0$ deg. Reflection intensities of $\alpha$-quartz $\rm I_{q}^{d}$, reflection intensities after applying the data processing method $\rm I_{q}^{c}$, mean background intensities $\rm \langle I_{b}\rangle$, background standard deviation average background intensity $\sigma_{b}$, values of the ``noise'' component of the background $\rm \Delta I_{b}$, the maximum intensity of pseudo-reflections after applying the data processing method $\rm I_{b-max}$ and the ratio of the reflection intensity of $\alpha$-quartz to the values of the background ``noise'' component $\rm I_{q}^{d}$ / $\rm \Delta I_{b}$  are presented in Table~\ref{tab7}.

%TABLE VII
\begin{table}[ht!]
\caption{Characteristics of reflections of $\alpha$-quartz $\rm I_{q}^{d}$  and background $\rm \langle I_{b}\rangle$ , $\rm \sigma_{b}$, ${\rm \Delta {I}_{b}}$; reflections of $\alpha$-quartz $\rm I_{q}^{c}$ and pseudo-reflections of background $\rm I_{b}^{c}$, obtained by data processing; the relationship $\rm I_{q}^{d}$/${\rm \Delta {I}_{b}}$ (counting time 60 - 5 s, angle interval $63.8-64.8$ degrees, background in the angle interval $61.0-63.0$ degrees, specimens of $\alpha$-quartz and LFCM brown ceramics are on a diffractometer).}
\centering
\begin{tabular} { r c c c c c c c c }

\hline
\hline 

 \strut No. & $\rm t,$ &  $\rm I_{q}^{d}$, & $\rm I_{q}^{c}$, & $\rm \langle I_{b}\rangle $, & $\rm \sigma_{b}$, & $\rm \Delta {I}_{b}$, & $\rm I_{b}^{c}$, & $\rm I_{q}^{d}$/$\rm \Delta {I}_{b}$,  \\

 { \enskip} & {s} & {counts} & {counts} & {counts} & {counts} & {counts} &{counts} & { \qquad}  \\
\hline

 \strut 1\,\,\, & 2 & 3 & 4 & 5 & 6 & 7 & 8 & 9  \\ 

\hline

{\strut 1.} & 60 & 6.8 & 4.0 & 13.1 & 2.4 & 9.6 & 2.8 &  0.71    \\

 2. & 55 & 6.2 & 4.2 & 13.4 & 2.6 & 10.4 & 3.3 & 0.60  \\

 3. & 50 & 5.6 & 2.8 & 12.8 & 3.1 & 12.4 & 1.9 & 0.45  \\

 4. & 45 & 5.1 & 4.8 & 14.0 & 3.4 & 13.6 & 1.6 & 0.38  \\
 
 5. & 40 & 4.5 & 3.1 & 13.8 & 3.8 & 15.2 & 4.7 & 0.30  \\
 
 6. & 35 & 4.0 & 4.4 & 13.1 & 3.2 & 12.8 & 2.7 & 0.31  \\
 
 7. & 30 & 3.4 & 3.1 & 13.5 & 4.1 & 16.4 & 3.7 & 0.21  \\
 
 8. & 25 & 2.8 & 3.4 & 12.6 & 3.7 & 14.8 & 1.4 & 0.19  \\
 
 9. & 20 & 2.3 & 4.6 & 13.4 & 5.6 & 22.4 & 4.6 & 0.10  \\
 
 10. & 15 & 1.7 & 4.8 & 12.5 & 4.8 & 19.2 & 5.5 & 0.09  \\
 
 11. & 10 & 1.1 & 5.5 & 13.7 & 6.6 & 26.4 & 5.8 & 0.04  \\
 
 12. & 5 & 0.6 & 4.7 & 15.6 & \!\!\!10.0 & 40.0 & 7.1 & 0.02  \\

\hline
\hline

\end{tabular}
\label{tab7}
\begin{flushleft}
\centering
	  \footnotesize{Note: The half-width of reflections ($\beta$) is equal to 0.3 deg;}  \\
    \end{flushleft}
\end{table}

The counting time during recording (Table~\ref{tab7}, column $2$) was gradually reduced from $60$ to $5$ s. The intensity of $\alpha$-quartz reflections $\rm I_{q}^{d}$ for a counting of $60$ s is taken from Table~\ref{tab6}, column $2$, row ``average''. For all smaller countings, the reflections' intensity is calculated proportionately to the shorter counting time. Naturally, the intensity of reflections (column $3$) decreases with decreasing counting time. The intensity of $\alpha$-quartz reflections after applying our data processing method $\rm I_{q}^{c}$ (column $4$) is practically independent of the counting time. The average values of the background intensities $\rm \langle I_{b}\rangle$ (column $5$), the standard deviation of the background $\rm \sigma_{b}$ (column $6$) and the values of the ``noise'' component of the background $\rm \Delta I_{b}$ (column $7$) show a slight tendency to increase with decreasing counting time. After applying the data processing method $\rm I_{q}^{c}$  (column $8$), the maximum intensity value of pseudo-reflections shows a larger increase with decreasing counting time. The ratio of the intensity of $\alpha$-quartz reflection to the values of the ``noise'' component of the background $\rm I_{q}^{d}$ / $\rm \Delta I_{b}$  (column $9$) decreases almost monotonically with decreasing counting time.

Analysis of the data in  Table~\ref{tab7} shows that for each counting time in the interval from $60$ to $45$ s, the intensities of $\alpha$-quartz reflection $\rm I_{q}^{d}$ (column $3$) are larger than the reflection intensities after applying the data processing method $\rm I_{q}^{c}$ (column $4$), and these reflection intensities ($\rm I_{q}^{c}$, column $4$) are in turn greater than the maximum values of the intensity of pseudo-reflections after applying the data processing method $\rm I_{b}^{c}$ (column $8$). In this counting time interval, the intensities of $\alpha$-quartz reflections on the diffraction and correlation patterns exceed the intensities of pseudo-reflections after applying our data processing method. This indicates that the method of processing X-ray diffraction data reliably reveals reflections of $\alpha$-quartz from the ``noise'' component of the background.

For each counting time in the interval from $40$ to $25$ s, the intensity of reflections of $\alpha$-quartz $\rm I_{q}^{d}$ (column $3$), the reflection intensity after applying the data processing method $\rm I_{q}^{c}$ (column $4$) and the maximum values of the intensity of pseudo-reflections after applying the processing method $\rm I_{b}^{c}$ (column $8$) have similar values. This indicates that at this counting time interval it is difficult to assess the efficiency of the X-ray diffraction data processing method for separating the $\alpha$-quartz reflection from the ``noise'' component of the background.

For each counting time in the interval from $20$ to $5$ s, the intensities of $\alpha$-quartz reflection $\rm I_{q}^{d}$ (column $3$) have lower values than the reflection intensities after applying our data processing method $\rm I_{q}^{c}$ (column $4$) and the maximum values of the intensity of pseudo-reflections after applying the data processing method $\rm I_{b}^{c}$ (column $8$). The intensity of reflections of $\alpha$-quartz on the diffraction pattern is less than the intensity of pseudo-reflections after applying the data processing method. This indicates that, in this counting interval, the method of processing X-ray diffraction data reveals $\alpha$-quartz reflections from the ``noise'' component of the background with low reliability.

Thus, the data in Table~\ref{tab7} allow us to conclude that at the ratio of the intensity of $\alpha$-quartz reflection to the values of the ``noise'' component of the background $\rm I_{q}^{d}$ / $\rm \Delta I_{b}$  up to $0.38$ (column $9$, counting time $45$ s), our method for processing of X-ray diffraction data makes it possible to reliably detect $\alpha$-quartz reflections from the ``noise'' component of the background. But when the ratio of the intensity of $\alpha$-quartz reflections to the values of the ``noise'' component of the background $\rm I_{q}^{d}$ / $\rm \Delta I_{b}$  is less than $0.10$ (column $9$, counting time $20$ s), the above method does not allow this. For cases where the ratio of the intensity of reflections of $\alpha$-quartz to the values of the ``noise'' component of the background $\rm I_{q}^{d}$ / $\rm \Delta I_{b}$ is in the range of $0.30-0.19$ (column $9$, counting time $40-25$ s), a deeper analysis of the data is required.

We plotted the intensity of $\alpha$-quartz reflections $\rm I_{q}^{d}$, the intensity of reflections after applying the data processing method $\rm I_{q}^{c}$ and the maximum intensity of pseudo-reflections after applying the data processing method ($\rm I_{b-max}$) versus counting time (Figre~\ref{f5}). Considering that after applying the data processing method, the reflection intensities $\rm I_{q}^{d}$ and the maximum values of the intensity of pseudo-reflections $\rm I_{q}^{c}$ have a significant spread in values due to significant errors in their measurement, we constructed best fit lines using the least square method for these intensities at time intervals of $60-25$ s and $20-5$ s.
% Figure 5.
\begin{figure}[ht!]
  \centering
  \vspace{1mm}
  \includegraphics[width=0.8\textwidth]{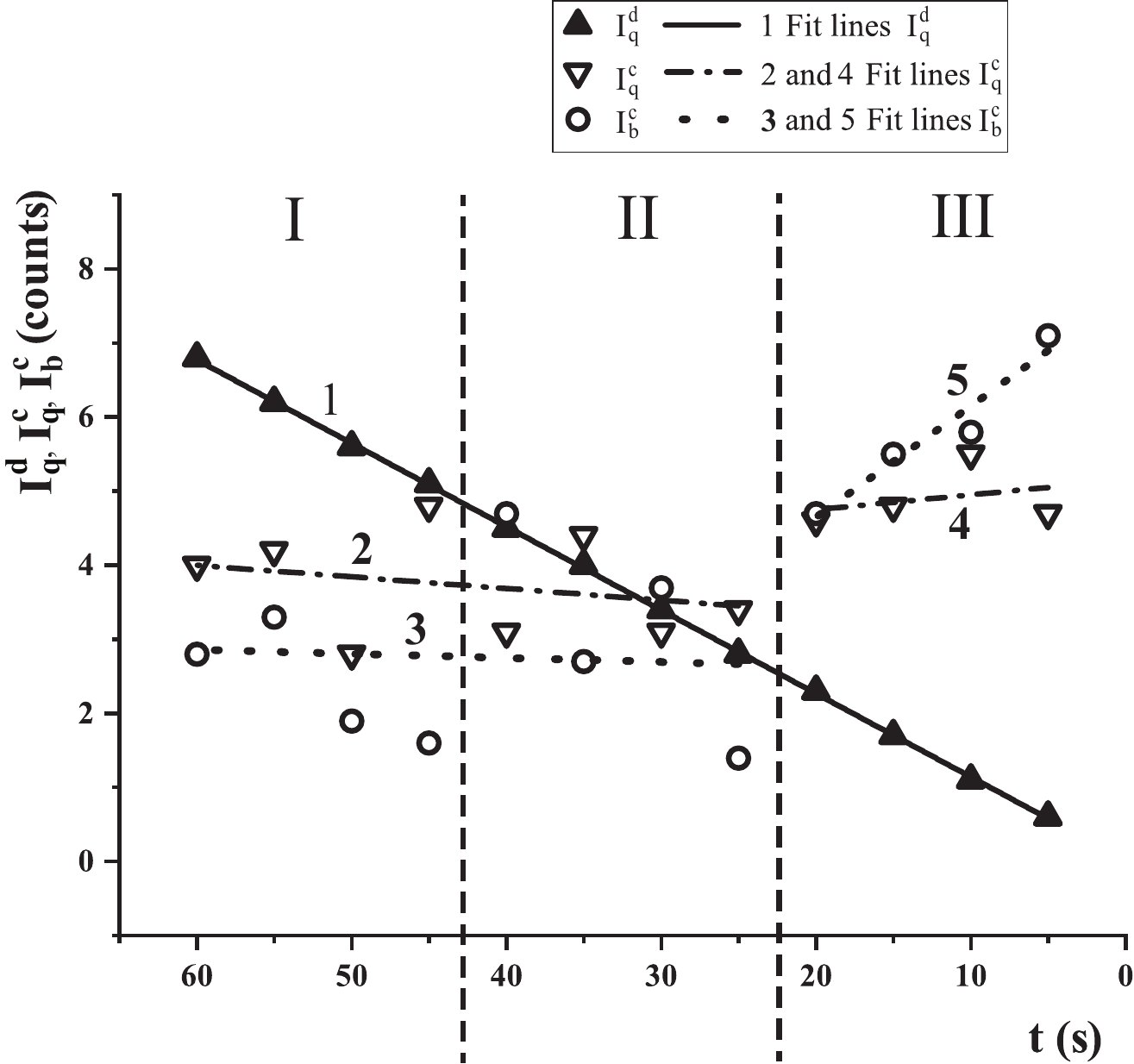}%
  \vspace{-1mm}
  \caption{The intensity of reflections of $\alpha$-quartz on the diffraction pattern $\rm I_{q}^{d}$, the intensity of the reflection on the correlation pattern after applying the data processing method $\rm I_{b}^{c}$, and the maximum values of the intensity of background reflections after applying the data processing method $\rm I_{b}^{c}$ versus exposure time. Fit lines: $\rm I_{q}^{d}$ -- 1, $\rm I_{q}^{c}$ -- 2 and 4 in the segments $60$--$25$ s and $20$--$5$ s, $\rm I_{b}^{c}$ -- 3 and 5 in the segments $60$--$25$ s and $20$--$5$ s, respectively.}
  \label{f5}
\end{figure}
Figure~\ref{f5} shows that for a counting time interval of $60-25$ s, the reflection intensities along the fit line after applying the data processing method ($\rm I_{q}^{c}$, straight line $2$) are greater than the maximum pseudo-reflection intensities after applying the data processing method ($\rm I_{b-max}$, straight line $3$). While for a counting time interval of $20-5$ s, the values along these linear splines change places (straight lines $4$ and $5$). This unambiguously indicates that in the range of counting times of $60-25$ s, the method of processing X-ray diffraction data makes it possible to detect $\alpha$-quartz reflections from the ``noise'' component of the background with high reliability, and in the range of counting times of $20-5$ s, it does not allow it.

Also, from the data in Figure~\ref{f5}, it follows that for a counting time interval of $40-25$ s, the values along the fit line of the $\alpha$-quartz reflection intensity $\rm I_{q}^{d}$ (straight line $1$) are comparable with the reflection intensity after applying the data processing method $\rm I_{q}^{c}$ (straight line $2$). This indicates a low reliability of the existence of reflections determined by the method of processing X-ray diffraction data in this interval of recording times, i.e. for the ratio of the intensity of $\alpha$-quartz reflection to the values of the ``noise'' component of the background $\rm I_{q}^{d}$ / $\rm \Delta I_{b}$  equal to $0.3-0.19$ (see Table~\ref{tab7}).

Figure~\ref{f6} shows segments of diffraction Figures~\ref{f6}(a),~\ref{f6}(c) and~\ref{f6}(f) and correlation Figures~\ref{f6}(b), ~\ref{f6}(d) and ~\ref{f6}(e) patterns of $\alpha$-quartz with brown ceramics. In Figure~\ref{f6}(a), with some difficulty, one can see the reflection of $\alpha$-quartz when its intensity is $0.6$ of the noise component of the background. In Figures~\ref{f6}(c) and Fig.~\ref{f6}(e), it is no longer possible to see the reflection of $\alpha$-quartz (the in-tensity of the reflection of $\alpha$-quartz is $0.38$ and $0.31$ of the values of the noise component of the background). Moreover, the X-ray diffraction data processing method makes it possible to identify these reflections with confidence (Figures~\ref{f6}(b), ~\ref{f6}(d) and ~\ref{f6}(f)).
\end{description}

\section{DISCUSSION AND ANALYSIS}

The practice of using X-ray phase analysis method shows that the vast majority of researchers usually obtain data at a high signal-to-noise ratio ~\citep{Guinebret}. If there is a need to work at a low signal-to-noise ratio, then much more often, one has to deal with reflections of low intensity due to a low content of crystalline phases with a significant number of these phases (up to eight or more). This situation occurs when we have studying the lava-like fuel-containing materials of the $4$th block of the Chornobyl NPP ~\citep{Gabielkov}. Much less often, one has to deal with the need to work with reflections of low intensity due to the time limitation for obtaining data on the phase composition of the material, study of steel during its heat treatment) ~\citep{Wiessner}.

The signal-to-noise ratio can be increased by decreasing the noise values. This can be achieved by using a particularly sensitive low noise detector ~\citep{Taguchi}. This method is more expensive than the method of processing X-ray diffraction data for multiphase materials with low phase content.

In the case of studying multiphase materials with a low content for some of phases, the signal-to-noise ratio can also be increased by increasing the signal values, i.e., increasing the recording time of one diffraction pattern. In order to ``pull out'' several reflections of low intensity for each of several crystalline phases, it will take time, calculated in tens of hours of diffractometer operation, which in most cases is unacceptable.

\begin{figure}[ht!]
  \centering
  \vspace{1mm}
  \includegraphics[width=0.8\textwidth]{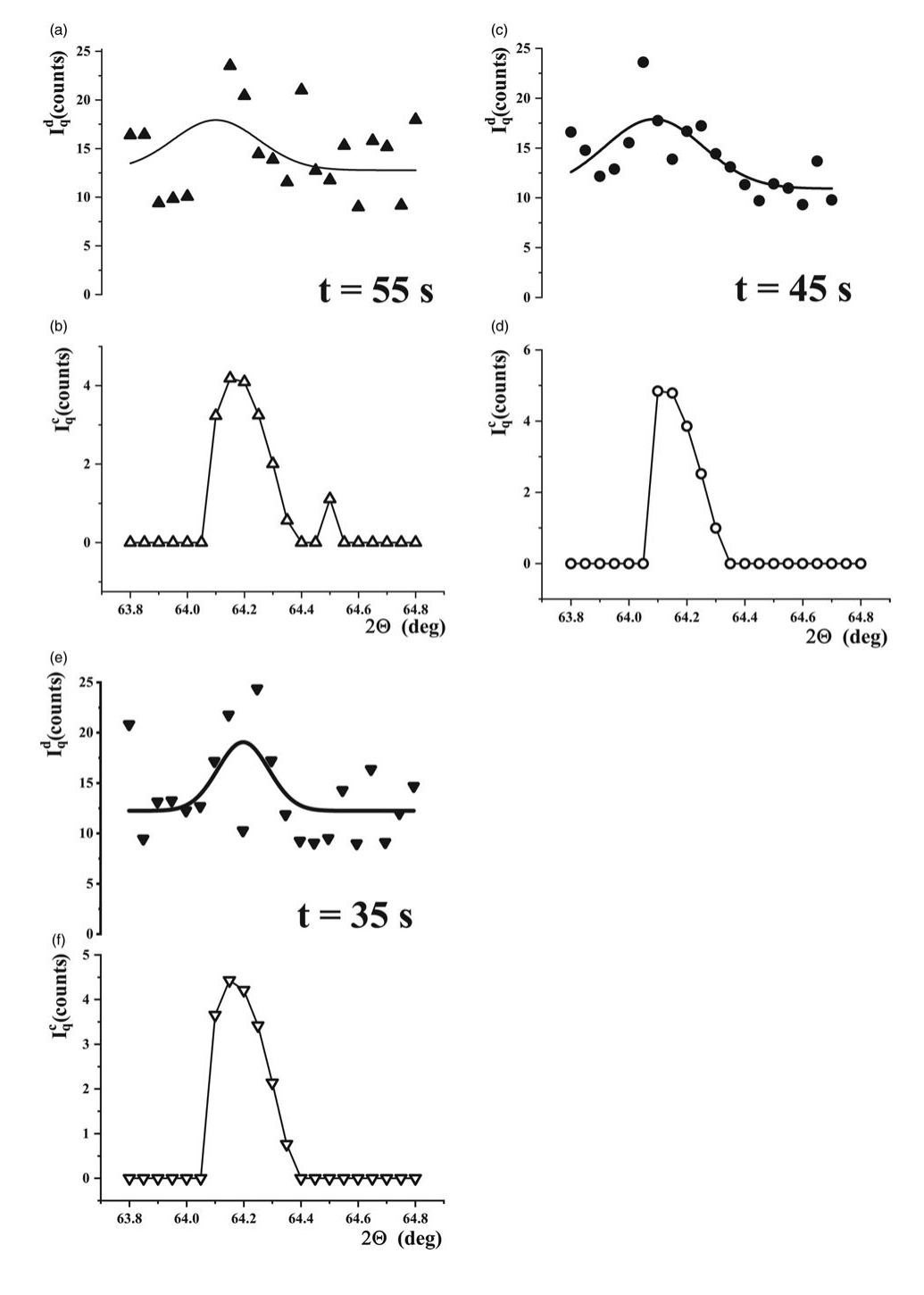}%
  \caption{Fragments of diffraction (Fig.~\ref{f6}(a), Fig.~\ref{f6}(c) and Fig.~\ref{f6}(f)) and correlation (Fig.~\ref{f6}(b), Fig.~\ref{f6}(d) and Fig.~\ref{f6}(e)) patterns of $\alpha$-quartz with brown ceramics (recording exposures: 55 s (Fig.~\ref{f6}(a) and Fig.~\ref{f6}(b)), 45 s (Fig.~\ref{f6}(c) and Fig.~\ref{f6}(d)), 35 s (Fig.~\ref{f6}(f) and Fig.~\ref{f6}(e)); signal-to-noise ratio: 0.6 (Fig.~\ref{f6}(a) and Fig.~\ref{f6}(b)), 0.38 (Fig.~\ref{f6}(c) and Fig.~\ref{f6}(d)), 0.31 (Fig.~\ref{f6}(f) and Fig.~\ref{f6}(e)).}
  \label{f6}
\end{figure}

The identification of weak reflections of crystalline phases with intensity less than the noise component of the background was achieved due to the use of the permutation test in the X-ray diffraction data processing method. We known about the use of the permutation test in the processing of X-ray diffraction data by French specialists ~\citep{Paradis-Fortin} in $2022$. We used a permutation test to detect weak reflections earlier in $2019$ ~\citep{Skorbun} in an article on our method for processing X-ray diffraction data for multiphase materials with a low content of phases.

To increase the background values, both the average values of the background intensity $\rm \langle I_{b}\rangle$ and the values of the ``noise'' component of the background $\rm \Delta I_{b}$, we used $\gamma$-quanta emitted by a sample of lava-like fuel-containing materials. Our series of experiments indicate that the following components contributed to the background recorded on the diffractometer: the background of the diffractometer registration system (instrument background), $\gamma$-quanta obtained as a result of beam scattering on a sample of $\alpha$-quartz and $\gamma$-quanta emitted by a sample of lava-like fuel-containing material.

Analysis of the data in Table~\ref{tab2},Table~\ref{tab3} and Table~\ref{tab4} shows that the average values of the background intensity $\rm \langle I_{b}\rangle$ (Table~\ref{tab4}, column $4$, counting time $195$ s) consist of $48$ \% from $\gamma$-quanta recorded by the diffractometer registration system, $18$ \% from $\gamma$-quanta scattered on $\alpha$-quartz and $34$ \% of $\gamma$-quanta emitted by a sample of lava-like fuel-containing material. The data in these tables also show that the ``noise'' component of the background $\rm \Delta I_{b}$ (Table~\ref{tab4}, column $6$, counting time $195$ s) consists of $80$ \% $\gamma$-quanta recorded by the diffractometer registration system and $20$ \% $\gamma$-quanta emitted by a sample of lava-like fuel-containing material. There is no contribution from $\gamma$-quanta scattered by the $\alpha$-quartz sample.

The results indicate that our method of processing X-ray diffraction data makes it possible to detect quartz reflections from the ``noise'' component of the background at the signal-to-noise ratio of more than $0.2$. It should be noted that at the signal-to-noise ratio in the range of $0.4-0.2$, the method of processing X-ray diffraction data gives less reliability. The reliability of the presence of reflections is low.

\section{CONCLUSION}
\begin{description}
\item[1.]Method of processing X-ray diffraction data makes it possible to reliably separate the reflections of crystalline phases at the signal-to-noise ratio of more than $0.4$. With a decrease in this ratio to $0.2$, the above method demonstrates less reliability, and the data obtained require a more detailed analysis. The method demonstrates practically unacceptable reliability at the signal-to-noise ratio of less than $0.2$, and the obtained data require confirmation.

\item[2.] The X-ray diffraction data processing method makes it possible to increase the possibilities of the X-ray phase analysis method due to the selection of reflections of crystalline phases, which are ``hidden'' in the noise of diffraction patterns. Using these two methods makes it possible to identify crystalline phases with a low (down to $0.1$ wt. \%) content. For most studied materials phases with such content are secondary, since they are caused by additives and inevitable impurities in the material (or in the original components from which the material under study is obtained). The possibility of identifying and evaluating the content of such phases will allow material developers to find out their role and degree of influence on the microstructure of the material under study and, accordingly, opens up opportunities for improving the physicochemical properties of materials.
\end{description}

\section{ACKNOWLEDGMENTS}
The work was sponsored in the framework of the budget theme of the National Academy of Sciences of Ukraine (No. $ \rm 0120U103480$).

\section{CONFLICT OF INTEREST}
The authors declare no conflicts of interest.

\end{document}